\documentclass[fleqn,twoside]{article}
\usepackage[headings]{espcrc2}

\readRCS
$Id: espcrc2.tex,v 1.2 2004/02/24 11:22:11 spepping Exp $
\usepackage{graphicx}
\usepackage[figuresright]{rotating}

\usepackage{epsfig,bbm,psfrag}
\usepackage{amsmath,amsfonts,amssymb}
\usepackage{citesort}
\usepackage[footnotesize]{caption}

\psfrag{tanb}{\small $\tan{\beta}$}
\psfrag{tan b}{\small $\tan{\beta}$}
\psfrag{log(mN)}{\small $\log(m_N)$}
\psfrag{BR(tau -> 3mu)}{\small BR$(\tau \to 3 \mu)$}
\psfrag{BR(tau -> 3e)}{\small BR$(\tau \to 3 e)$}
\psfrag{BR(mu -> 3e)}{\small BR$(\mu \to 3 e)$}
\psfrag{BR(mu ->3e)}{\small BR$(\mu \to 3 e)$}
\psfrag{BR(tau -> mu + gamma)}{\small BR$(\tau \to \mu \gamma)$}
\psfrag{BR(tau -> e + gamma)}{\small BR$(\tau \to e \gamma)$}
\psfrag{BR(mu -> e + gamma)}{\small BR$(\mu \to e \gamma)$}
\psfrag{mod(theta2)}{\small $|\theta_2|$}
\psfrag{mod(theta1)}{\small $|\theta_1|$}

\title{LFV in tau and muon decays within SUSY seesaw}

\author{S. Antusch\address[FTUAM]{Departamento de F\'{\i }sica Te\'{o}rica C-XI 
and Instituto de F\'{\i }sica Te\'{o}rica C-XVI, \\
	Universidad Aut\'{o}noma de Madrid,
	Cantoblanco, E-28049 Madrid, Spain},
        E. Arganda\addressmark[FTUAM],
        M.J. Herrero\addressmark[FTUAM]\thanks{Talk given at the 9th International Workshop on Tau-Lepton Physics, Tau06, 19-22 September 2006, Pisa (Italy)}
        and
        A.M. Teixeira\addressmark[FTUAM]}

\runtitle{LFV in tau and muon decays within SUSY seesaw}
\runauthor{S. Antusch, E. Arganda, M.J. Herrero and A.M. Teixeira}

\begin{document}

\begin{abstract}
In these proceedings we present the results for lepton flavour violating
tau and muon decays within
the SUSY seesaw scenario. Specifically, we consider the Constrained 
Minimal Supersymmetric Standard Model extended by 
three right handed neutrinos, $\nu_{R_i}$ and their corresponding 
SUSY partners, ${\tilde \nu}_{R_i}$, ($i=1,2,3$),
and use the seesaw mechanism for neutrino mass generation. We include   
the predictions for the branching ratios of two types of lepton flavour violating channels, 
$l_j \to \l_i \gamma$
and $l_j \to 3 l_i$, and compare them with the present bounds and future
experimental sensitivities. We first analyse the
dependence of the branching ratios with the most relevant SUSY seesaw
parameters, and we then focus 
on the particular 
sensitivity to $\theta_{13}$, which we find specially interesting on the light of its
potential future
measurement. We further study the constraints from the requirement of successfully 
producing the
baryon asymmetry of the Universe via thermal leptogenesis, which is another 
appealing
feature of the SUSY seesaw scenario.
We conclude with the impact that a potential measurement of $\theta_{13}$ can
have on lepton flavour violating physics.    
This is a very short summary of the works in Refs.~\cite{Arganda:2005ji} and~\cite{Antusch:2006vw} to which we
refer the reader for more details.
\vspace{1pc}
\end{abstract}

\maketitle

\section{LFV within SUSY seesaw}

The seesaw mechanism
is implemented by the inclusion of a Majorana mass $m_R$ for the right handed
neutrinos (allowed due to their singlet character under all the 
symmetries of the Standard Model (SM)) and by considering a large separation between this mass
and the
electroweak (EW) scale~\cite{seesaw:I}. After EW symmetry breaking, the full $6 \times 6$ 
neutrino 
mass matrix is given in terms of the 
$3 \times 3$ Dirac mass
matrix, $m_D=Y_\nu \langle H_2 \rangle$,  and the $3 \times 3$ Majorana mass matrix $m_R$. 
Here $Y_\nu$ is the $3 \times 3$ neutrino Yukawa coupling and 
$\langle H_2 \rangle=v \sin \beta$ with $v=174$ GeV. The ratio of the two
Higgs doublets vacuum expectations values
is $\tan \beta= \langle H_2 \rangle/\langle H_1 \rangle$.     
The
assumption of $v \ll m_R$ leads to
the usual seesaw equation, $m_\nu=-m_D^Tm_R^{-1}m_D$, which guaranties the
smallness of the light neutrino masses. 
After the diagonalisation of the full 
$6 \times 6$ 
neutrino 
mass matrix one obtains six physical Majorana neutrinos: 
three light $\nu_i$, with masses $m_\nu^{\rm diag}={\rm
diag}(m_{\nu_1},m_{\nu_2},m_{\nu_3})= U_\text{MNS}^T m_{\nu}U_\text{MNS}$, 
and three heavy $N_i$, with masses $m_N^{\rm diag}={\rm
diag}(m_{N_1},m_{N_2},m_{N_3})=m_R$. Notice that we work in a lepton basis where 
both the right handed
mass matrix and the charged lepton mass matrix are diagonal in flavour space. 
The flavour mixing in the light neutrino sector is given by 
the Maki-Nakagawa-Sakata matrix  $U_\text{MNS}$~\cite{Umns} 
for which we use the standard parameterization, which is written in terms of three mixing angles 
$\theta_{12}$, $\theta_{13}$ and $\theta_{23}$ and three CP violating phases $\delta$,
$\varphi_1$ and $\varphi_2$.

We use here the parameterisation proposed in Ref.~\cite{Casas:2001sr}, where
the solution to the seesaw equation is written as 
$m_D = \sqrt {m_N^{\rm diag}} R \sqrt {m_\nu^{\rm diag}}U^{\dagger}_\text{MNS}$, 
with $R$ being a $3 \times
3$ orthogonal complex matrix, defined by three complex angles $\theta_i$
($i=1,2,3$). The attractiveness of this parameterisation is that it allows to
easily implement the requirement of compatibility with low energy neutrino
data. It also clearly shows that
in the singlet seesaw scenario one can have large neutrino Yukawa couplings, $Y_{\nu} \sim {\cal
O}(1)$,   by simply choosing large
entries in $m_N^{\rm diag}$. 
The main implication of these large Yukawa coupling is that they can induce
large lepton flavour violating (LFV) rates~\cite{Borzumati:1986qx}.
The total number of
parameters of the neutrino sector in this scenario is 18, which in this 
particular parameterisation are summarised by $\theta_{ij}$, $\delta$, $\varphi_1$, 
$\varphi_2$, $m_{\nu_i}$, $m_{N_i}$ and $\theta_i$. By adjusting the light neutrino 
parameters to the low energy neutrino data, one is left with 9 input parameters given by 
$m_{N_i}$ and $\theta_i$. 

Regarding the numerical estimates we consider two scenarios. The first one with quasi-degenerate
 light neutrinos, with masses 
 $m_{\nu_1}=0.2$ eV, $m_{\nu_2}=m_{\nu_1}+\frac{\Delta m_{sol}^2}{2 m_{\nu_1}}$ and 
 $m_{\nu_3}=m_{\nu_1}+\frac{\Delta m_{atm}^2}{2 m_{\nu_1}}$, and degenerate heavy neutrinos
 with mass $m_N$. The second one is with hierarchical light and heavy neutrinos, with
 masses  
 $m_{\nu_1} \, \ll \,   m_{\nu_2} = \sqrt{{\Delta
m_{sol}^2}} \, \ll \, m_{\nu_3} = \sqrt{{\Delta m_{atm}^2}},$  and  
$m_{N_1} \ll  m_{N_2} \ll m_{N_3}$. Here we use 
$\sqrt{\Delta m_\text{sol}^2}=0.009$ eV, $\sqrt{\Delta m_\text{atm}^2}=0.05$ eV, 
$\theta_{12}=\theta_\text{sol}=30^\circ$, $\theta_{23}=\theta_\text{atm}=45^\circ$, 
$0^\circ\leq \theta_{13} \leq 10^\circ$, and for simplicity we fix 
$\delta = \varphi_1=\varphi_2=0$.
 
In addition to the previous seesaw parameters, there are the SUSY
sector parameters which, within the assumed
Constrained Minimal Supersymmetric Standard Model (CMSSM) framework, are given by $M_{1/2},\,M_0,\,A_0,\,\tan\beta,$ and 
$\text{sign}\,\mu$. The universality of the soft-SUSY breaking terms is imposed at a high
scale $M_X$
which we fix here to the $g_2$-$g_1$ gauge couplings unification scale $M_X=2 \times 10^{16}$ GeV. In particular, in the numerical analysis we consider specific
choices of these parameters, given by the mSUGRA-like ``Snowmass Points and Slopes''
(SPS) \cite{Allanach:2002nj} listed in Table~\ref{SPS:def:15}, which represent different examples of possible 
SUSY spectra. 

\begin{table*}
\begin{center}
\begin{tabular}{|c|c|c|c|c|c|}
\hline
SPS & $M_{1/2}$ (GeV) & $M_0$ (GeV) & $A_0$ (GeV) & $\tan \beta$ & 
 $\mu$ \\\hline
 1\,a & 250 & 100 & -100 & 10 &  $>\,0 $ \\
 1\,b & 400 & 200 & 0 & 30 &   $>\,0 $ \\
 2 &  300 & 1450 & 0 & 10 &  $>\,0 $ \\
 3 &  400 & 90 & 0 & 10 &    $>\,0 $\\
 4 &  300 & 400 & 0 & 50 &   $>\,0 $ \\
 5 &  300 & 150 & -1000 & 5 &   $>\,0 $\\\hline
\end{tabular} 
\caption{Values of $M_{1/2}$, $M_0$, $A_0$, $\tan \beta$, 
and sign $\mu$ for the SPS points considered in the analysis.}
\label{SPS:def:15}
\end{center}
\end{table*}

Regarding the technical aspects of the computation of the branching ratios, they are 
explained in detail in Refs.~\cite{Arganda:2005ji} and~\cite{Antusch:2006vw}. Here we only summarise the most relevant points: 
\begin{itemize}
\item[{\large $\bullet$}] It is a full-one loop computation of branching
  ratios (BRs), i.e., we include all
contributing one-loop diagrams with the SUSY particles flowing in the loops. For the case 
of $\l_j \to l_i \gamma$ the analytical formulae can be found in~\cite{Hisano:1995cp,Arganda:2005ji,Hisano:1995cp}. For the
case $\l_j \to 3 \l_i$ the complete set of diagrams (including photon-penguin, $Z$-penguin,
Higgs-penguin and box diagrams) and formulae are given in~\cite{Arganda:2005ji}.   
\item[{\large $\bullet$}] The computation is performed in the physical basis for all SUSY
particles entering in the loops. In other words, we do not
use the Mass Insertion Approximation (MIA). 
\item[{\large $\bullet$}] The running of the CMSSM-seesaw parameters from the
universal scale $M_X$ down to the electroweak scale is performed by
numerically solving the full one-loop Renormalisation Group Equations (RGEs) (including the extended neutrino sector) 
and by means of the public Fortran Code
SPheno2.2.2.~\cite{Porod:2003um}. More concretely, we do not use the Leading Log Approximation (LLog).
\item[{\large $\bullet$}] The light neutrino sector parameters that are used in 
$m_D = \sqrt {m_N^{\rm diag}} R \sqrt {m_\nu^{\rm diag}}U^{\dagger}_\text{MNS}$
are those evaluated at the seesaw scale $m_R$. That is, we start with their low energy values 
(taken from data) and then apply the RGEs to run them up to $m_R$. 
\item[{\large $\bullet$}]
We have added to the SPheno code extra subroutines that compute the LFV
rates for all the $\l_j \to l_i \gamma$ and $\l_j \to 3 \l_i$ channels. 
We have also included additional subroutines to: implement 
the requirement of succesfull baryon asymmetry of the Universe (BAU), wich we define as having  
$n_B/n_\gamma \in [10^{-10},10^{-9}]$; implement the requirement of compability with present bounds on lepton
electric dipole moments: {$\mbox{EDM}_{e \mu \tau}$} {$\lesssim (6.9
\times 10^{-28}, 3.7 \times 10^{-19}, 4.5 \times 10^{-17}) \, \mbox{e.cm}$}.  
\end{itemize}

In what follows we present the main results for degenerate and hierarchical
heavy neutrinos. We also include a
comparison with present bounds on LFV rates~\cite{Brooks:1999pu,Aubert:2005wa,Aubert:2005ye,Bellgardt:1987du,Aubert:2003pc} and their future sensitivities~\cite{mue:Ritt,Akeroyd:2004mj,Iijima,Aysto:2001zs,PRIME,Kuno:2005mm}.

\section{Results for degenerate heavy neutrinos}

\begin{figure*}[t]
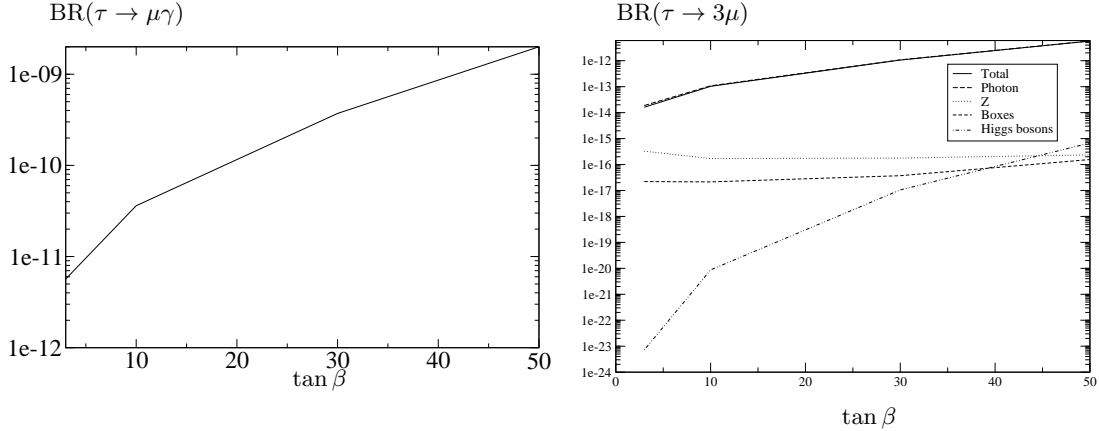

  \begin{center} \hspace*{-10mm}
    \begin{tabular}{cc}
      \psfig{file=BRtaumugamma_tanb_deg.epsi,width=50mm,angle=270,clip=}
      &
      \psfig{file=BRtau3mu_tanb_deg.epsi,width=55mm,angle=270,clip=}
    \end{tabular}
    \caption{Predictions for LFV tau decay rates for degenerate heavy neutrinos as a function 
    of $\tan\beta$. The remaining SUSY parameters are as for SPS4. The seesaw parameters are
     $m_N=10^{14}$ GeV and $R$ is any real orthogonal matrix. Here we have set
     $\theta_{13} = 0^\circ$} 
    \label{fig:degenerate}
  \end{center}
\end{figure*}

In this case, the most relevant parameters are the common heavy mass
$m_N$ and $\tan \beta$. Notice that by choosing a real $R$-matrix the rates do
not depend on the
particular value of the $R$-matrix entries.
The alternative case of a complex $R$-matrix for degenerate neutrinos has been
analysed in~\cite{Pascoli:2003rq} and leads in general to larger LFV rates.
We have found that for both LFV processes
$\l_j \to l_i \gamma$  and $l_j \to 3 l_i$, the full BRs grow with $m_N$ approximately
as $(m_N \log m_N)^2$, in agreement with what is expected from the LLog
approximation. We have explored here the range 
$10^8 \, {\rm GeV} \leq  m_N  \leq 10^{14}$ GeV.
Therefore the maximun rates found are associated with the largest considered value of 
$m_N=10^{14}$ GeV. Regarding $\tan \beta$, we have found that both rates 
BR$(l_j \to l_i \gamma)$ and BR$(l_j \to 3 l_i)$ grow 
approximately as $\tan^2 \beta$, as is expected in the MIA. This can be
clearly seen
in Fig.~\ref{fig:degenerate} where 
the BR predictions for the channels with largest rates, $\tau \to \mu \gamma$ and
$\tau \to 3 \mu$, are shown as a 
function of $\tan \beta$. It is also manifest from this figure that the dominant
contribution to BR$(\tau \to 3 \mu)$, by many orders of magnitude, comes from the
photon-penguin diagrams (superimposed on the total in this figure), even at
large $\tan \beta$, 
where the Higgs-penguin contributions
get their maximum values. This demonstrates that the approximate formula,
BR$(l_j \to 3 l_i)/$BR$(l_j \to l_i \gamma)=
\frac{\alpha}{3\pi}\left(\log\frac{m_{l_j}^2}{m_{l_i}^2}-\frac{11}{4}\right)$, 
 leading to values 
of $\frac{1}{440}$, $\frac{1}{94}$ and $\frac{1}{162}$ respectively for 
 $(l_jl_i)= (\tau \mu), (\tau e)$ and  $(\mu e)$, works extremely
 well.
 
In summary, for the explored parameters range, we have found LFV rates that, 
are all bellow the present upper experimental  
bounds. The largest ratios found are for $m_N=10^{14}\, {\rm GeV}$ and
$\tan\beta= 50$.  For instance, by choosing the SPS4 point we obtain
BR$(\tau \to 3 \mu)^{\rm max} \sim  3 \times 10^{-11}$ and
BR$(\tau \to \mu \gamma)^{\rm max} \sim 10^{-8}$. Regarding the other SPS points, 
and for a given choice of the seesaw
parameters, we find quite generically the following hierarchy among the corresponding 
BRs: BR$_{\rm SPS4}$~$>$ BR$_{\rm SPS1b}$~$\gtrsim$ BR$_{\rm SPS1a}$
$>$ BR$_{\rm SPS3}$~$\gtrsim$ BR$_{\rm SPS2}$~$>$ BR$_{\rm SPS5}$.

\section{Results for hierachical heavy neutrinos}

We next present the results for the alternative case of hierarchical heavy
neutrinos where we find rates that are, for some regions of the SUSY-seesaw 
parameter space, 
within the present and/or future experimental reach. 
In this case, the BRs are mostly sensitive to the heaviest
mass $m_{N_3}$, $\tan\beta$, $\theta_1$ and $\theta_2$. The other input seesaw
parameters $m_{N_1}$, $m_{N_2}$ and $\theta_3$ play a secondary role since the BRs do
not strongly depend on them. The dependence on $m_{N_1}$ and $\theta_3$ appears only 
indirectly, once the requirement of a successfull BAU is imposed. We will comment
more on this later. 

\begin{figure*}[t]
  \begin{center} \hspace*{-10mm}
    \begin{tabular}{cc}
       \psfig{file=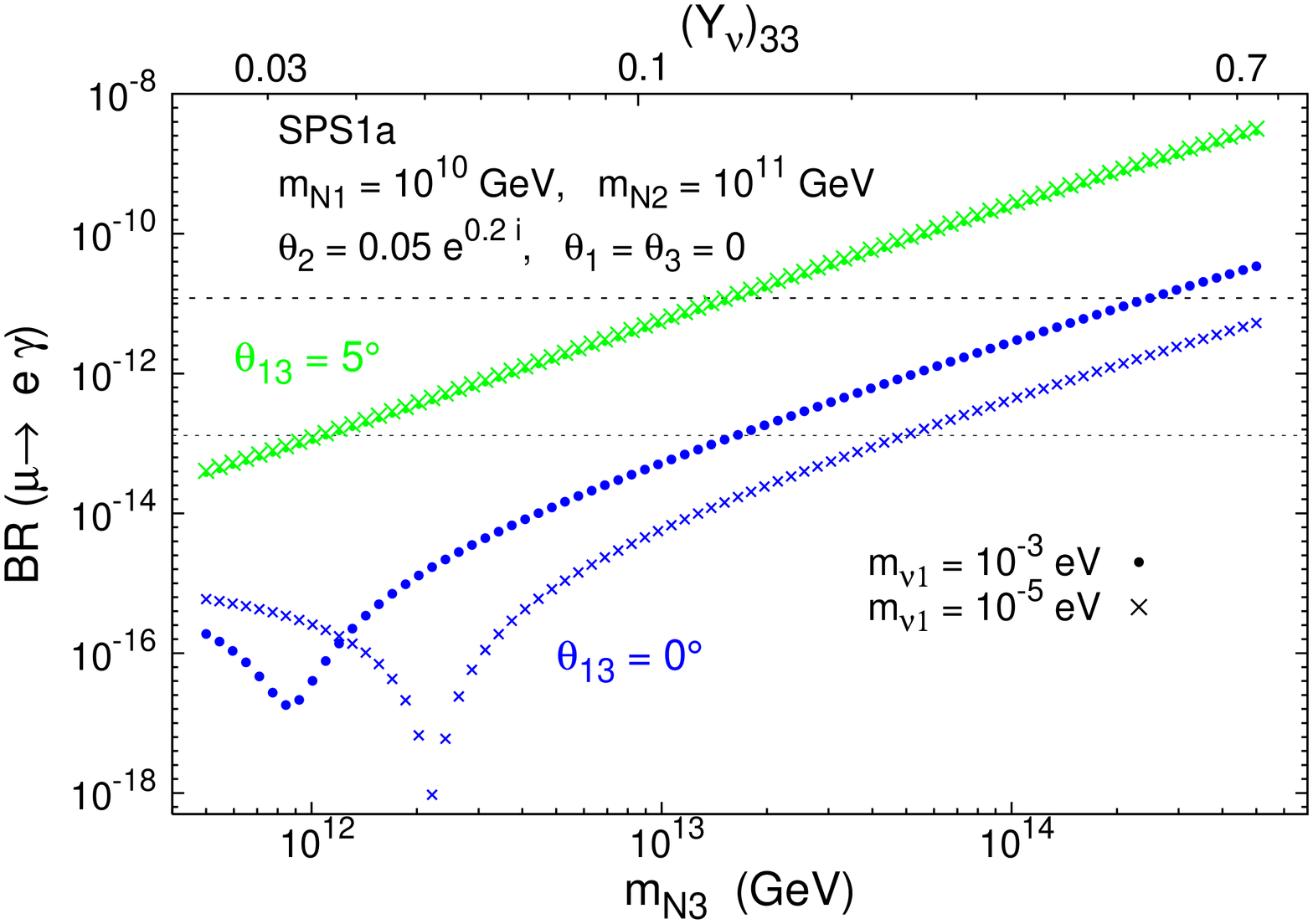,width=55mm,angle=270,clip=}&
       \psfig{file=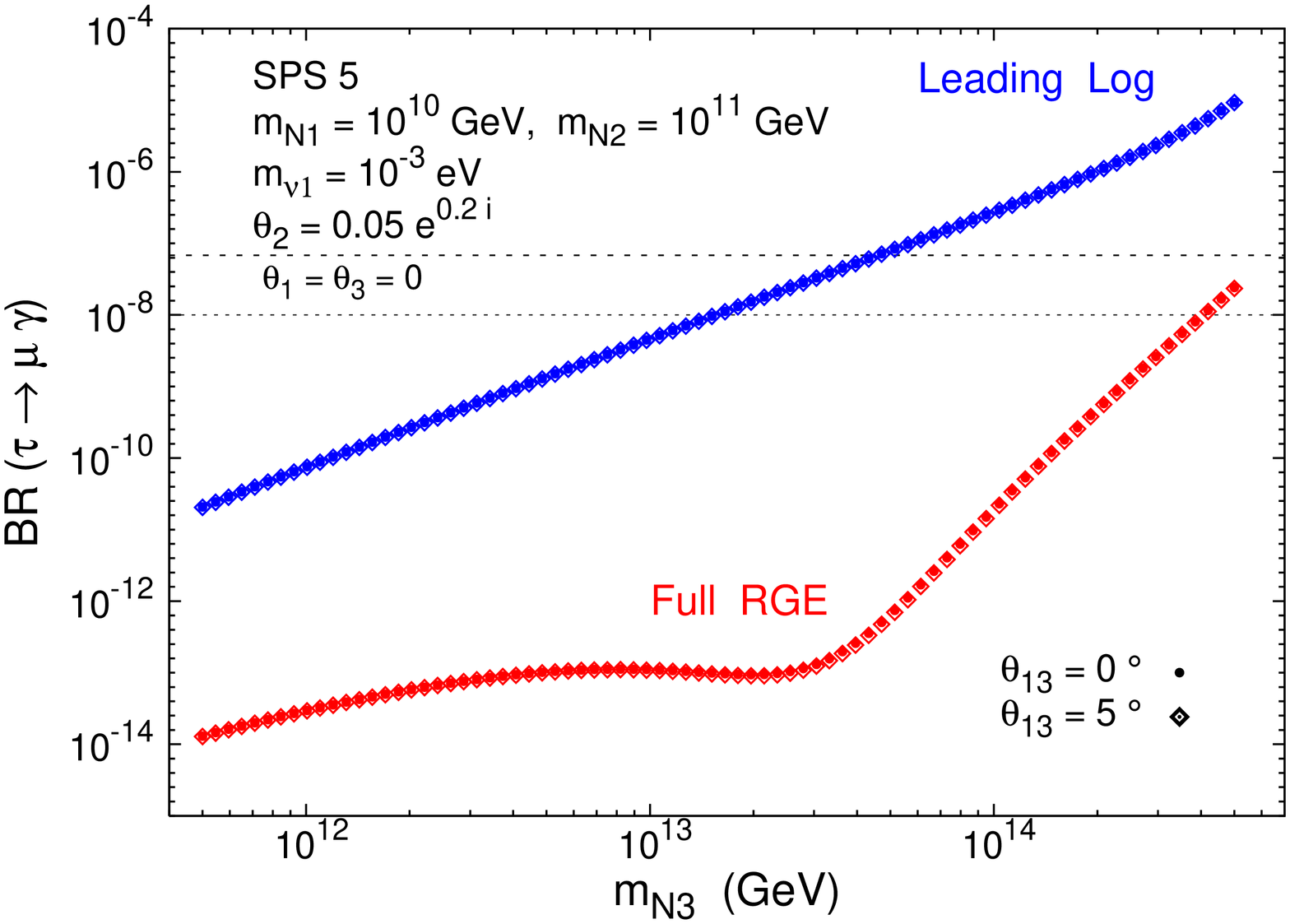,width=55mm,angle=270,clip=}
    \end{tabular}
  \caption{On the left, BR($\mu \to e\, \gamma$) as a function of 
      $m_{N_3}$ for SPS 1a, with
      $m_{\nu_1}\,=\,10^{-5}$~eV and $m_{\nu_1}\,=\,10^{-3}$~eV (times, dots,
      respectively), and
      $\theta_{13}=0^\circ,\,5^\circ$ (blue/darker, green/lighter lines). 
      Baryogenesis is enabled by the choice 
      $\theta_2\,=0.05\,e^{0.2\,i}$ ($\theta_1=\theta_3=0$). 
      On the upper horizontal axis we display the associated value of
      $(Y_\nu)_{33}$.
      On the right, BR($\tau \to \mu \, \gamma$) as a function of 
      $m_{N_3}$ for SPS5, with
      $m_{\nu_1}\,=\,10^{-3}$~eV and $\theta_2\,=0.05\,e^{0.2\,i}$
      ($\theta_1 = \theta_3 = 0^\circ$).
      In both cases a dashed
      (dotted) horizontal line denotes the present experimental bound (future sensitivity).}  
    \label{fig:MN3:MN2:SPS1a}
  \end{center}
\end{figure*}
\begin{figure*}
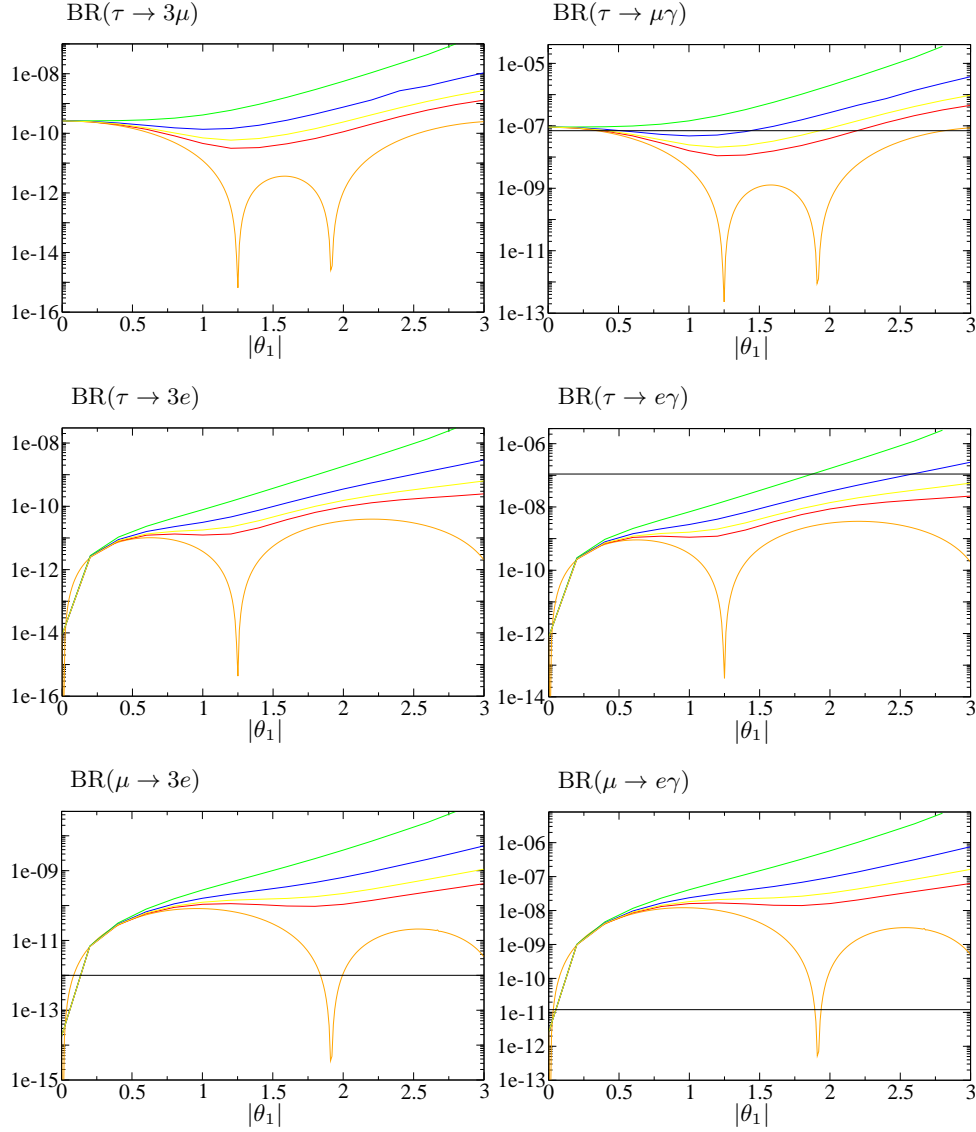

\vspace{-1.5cm}
\begin{center}
\psfig{figure=BRtau3mu_argtheta1.epsi,height=2.5in,angle=-90}
\psfig{figure=BRtaumugamma_argtheta1.epsi,height=2.5in,angle=-90}\\
\vspace{0.5cm}
\psfig{figure=BRtau3e_argtheta1.epsi,height=2.5in,angle=-90}
\psfig{figure=BRtauegamma_argtheta1.epsi,height=2.5in,angle=-90}\\
\vspace{0.5cm}
\psfig{figure=BRmu3e_argtheta1.epsi,height=2.5in,angle=-90}
\psfig{figure=BRmuegamma_argtheta1.epsi ,height=2.5in,angle=-90}
\end{center}
\caption{Predictions of LFV $\tau$ and $\mu$ decay rates as a function of 
$\vert \theta_1 \vert$ for hierarchical heavy neutrinos, complex $R$-matrix,
and for SPS4. 
The seesaw parameters are $\arg(\theta_1) = 0, \pi/10, \pi/8, \pi/6, \pi/4$ 
(lower to upper lines), $\theta_2 = \theta_2 = 0$ and the heavy neutrino masses are $(m_{N_1},m_{N_2},m_{N_3})=(10^8, 
2 \times 10^8, 10^{14})$ GeV. The horizontal lines denote 
the present experimental bounds. Here we have set $\theta_{13} = 0^\circ$}.
\label{fig:modth1}
\end{figure*} 

We display in Fig.~\ref{fig:MN3:MN2:SPS1a} the predictions for BR$(\mu \to e \gamma)$ and BR$(\tau \to \mu
\gamma)$ as a function of $m_{N_3}$, for a specific choice of the other
input parameters. This figure clearly shows the strong sensitivity of the BRs 
to $m_{N_3}$. In fact, the BRs vary by as much as six orders of magnitude in the explored range
of $5 \times 10^{11} \, {\rm GeV} \leq  m_{N_3} \leq 5 \times 10^{14} \, {\rm GeV}$.
Notice also that for the largest values of $m_{N_3}$ considered, 
the predicted rates for $\mu \to e \gamma$ are within the present experimental
reach while those of $\tau \to \mu \gamma$ are only within the future experimental sensitivity. 
It is also worth mentioning 
that by
comparing our full results with the LLog predictions, we find that the LLog
approximation dramaticaly fails in some cases. In particular, for the SPS5 point, the 
LLog predictions overestimate the BRs by about four orders of magnitude. For
the other points
SPS4, SPS1a,b and SPS2 the LLog estimate is very similar to the full result, whereas 
for SPS3 it underestimates the full computation by a factor of three. 
In general, the divergence 
of the LLog and the full computation occurs for low $M_0$ and large $M_{1/2}$~\cite{Petcov:2003zb,Chankowski:2004jc} and/or large
$A_0$ values~\cite{Antusch:2006vw}. The failure of the LLog is more dramatic for SUSY scenarios with
 large $A_0$. 
Fig.~\ref{fig:MN3:MN2:SPS1a}
also shows that while in some cases (for instance SPS1a)
the behaviour of the BR with $m_{N_3}$ does 
follow the expected LLog approximation (BR $\sim (m_{N_3}\log m_{N_3})^2$), there
are other scenarios where this is not the case. A good example of this is SPS5.
It is also worth commenting on the deep minima of BR$(\mu \to e \gamma)$ appearing in
Fig.~\ref{fig:MN3:MN2:SPS1a} for the lines
associated with $\theta_{13}=0^\circ$.  These minima
are induced by the effect of the running of $\theta_{13}$, shifting it
from zero to a negative value (or equivantly $\theta_{13} > 0$ and $\delta = \pi$).
In the LLog approximation, 
they  can be understood  as
a cancellation occurring in the relevant matrix element of $Y_\nu^{\dagger} L Y_\nu$, with $L_{ij}=
\log (M_X/m_{N_i})\delta_{ij}$. Explicitly, the cancellation occurs between the terms proportional to $m_{N_3}\,L_{33}$ and 
$m_{N_2}\,L_{22}$ in the limit $\theta_{13}(m_R) \to 0^-$ 
(with $\theta_1=\theta_3=0$).
The depth of these minima is larger for smaller $m_{\nu_1}$,
as is visible in Fig.~\ref{fig:MN3:MN2:SPS1a}.

Regarding the $\tan \beta$ dependence of the BRs we obtain that, similar to
what was found for the degenerate case, the BR grow as $\tan^2 \beta$. The
hierarchy of the BR predictions for the several SPS points is dictated by the
corresponding $\tan \beta$ value, with a secondary role being played by the
given SUSY spectra. We again find the following generic hierarchy:
BR$_{\rm SPS4}$~$>$ BR$_{\rm SPS1b}$~$\gtrsim$ BR$_{\rm SPS1a}$ 
$>$ BR$_{\rm SPS3}$~$\gtrsim$ BR$_{\rm SPS2}$~$>$ BR$_{\rm SPS5}$.

In what concerns to the $\theta_i$ dependence of the BRs, we have found that they are 
mostly sensitive to $\theta_1$ and $\theta_2$. The BRs are nearly
constant with $\theta_3$. The predictions for all the LFV channels
as functions of $\theta_1$ are shown in Fig.~\ref{fig:modth1}. From this
figure we first see that the BRs basically follow the pattern of the $Y_\nu$ couplings as functions of
$\theta_1$, including the
appearance of pronounced dips at particular $|\theta_1|$ values for the real 
$\theta_1$ case. Although not displayed here, the results for $Y_\nu$ 
show that the largest predicted entries are $Y_\nu^{33}$ and $Y_\nu^{32}$, 
reaching values up to ${\cal O}(1)$ for the explored $\theta_1$ range 
(see also the left panel of Fig.~\ref{fig:MN3:MN2:SPS1a}).   
The main conclusion from Fig.~\ref{fig:modth1} is that
the predictions for BR$(\mu \to e \gamma)$, BR$(\mu \to 3 e)$, BR$(\tau \to \mu
\gamma)$ and BR$(\tau \to e \gamma)$ are above their corresponding experimental
bound for specific values of $\theta_1$. Particularly, the LFV muon decay rates are 
well above their present experimental bounds for most of the
explored $\theta_1$ values. Notice also that for SPS4 the predicted BR$(\tau \to \mu
\gamma)$ values are very close to the present experimental reach even at
$\theta_1=0$ (that is, $R=1$). We have also explored the dependence on $\theta_2$ and
found similar results (not shown here), with the appearance of pronounced dips
at particular real values of
$\theta_2$ with the BR$(\mu \to e \gamma)$, BR$(\mu \to 3 e)$ and BR$(\tau \to \mu
\gamma)$ predictions being above the experimental bounds for some $\theta_2$ values.

We next address the sensitivity of the LFV BRs to $\theta_{13}$. We first present the 
results for the case $R=1$ and then discuss how this sensitivity
changes when
moving from this case towards the more general case of a complex $R$, taking
into account additional constraints from the requirement of a succesfull BAU. 
 
\begin{figure*}[t]
  \begin{center} \hspace*{-10mm}
    \begin{tabular}{cc}
	\psfig{file=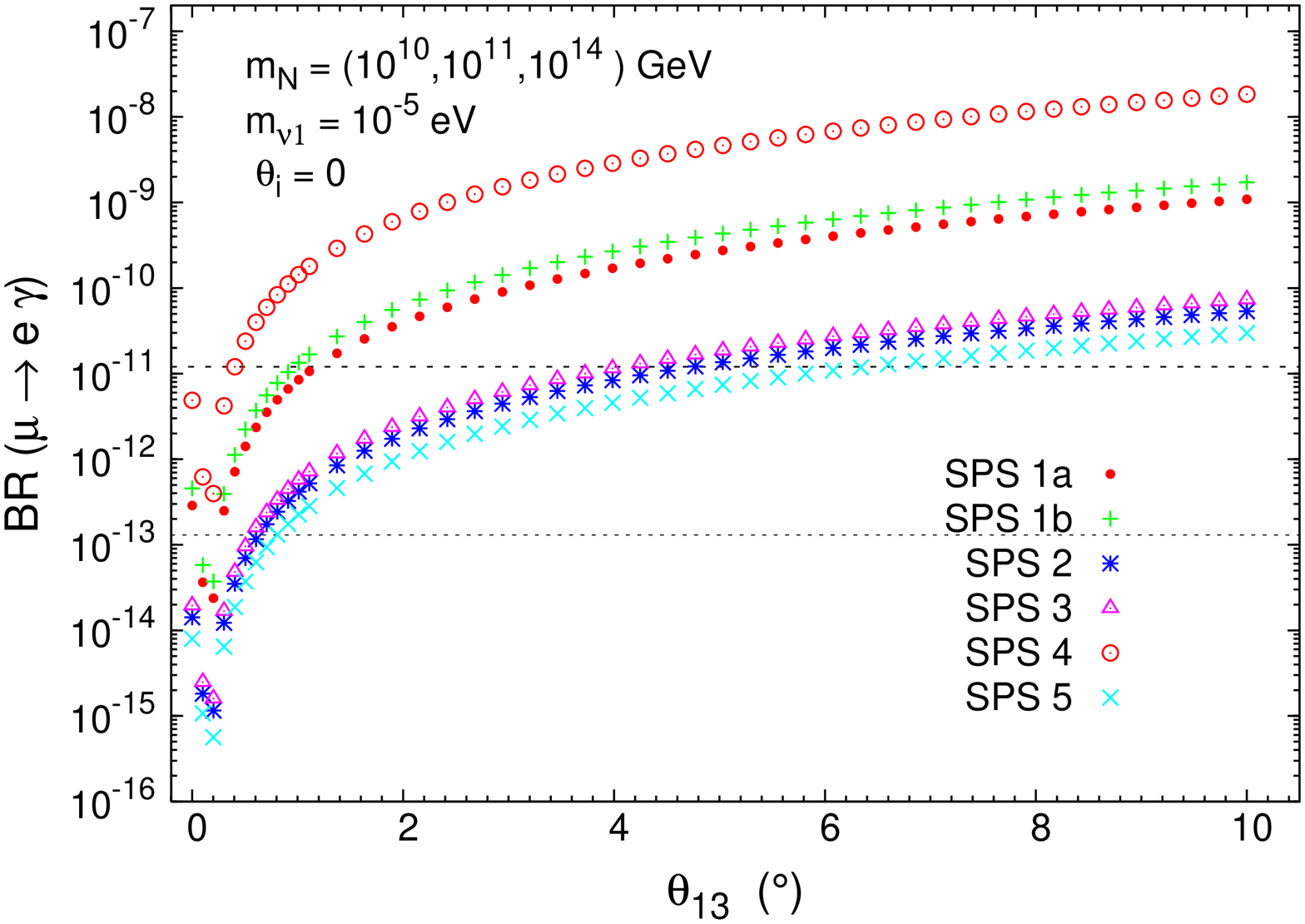,width=50mm,angle=270,clip=} &
        \psfig{file=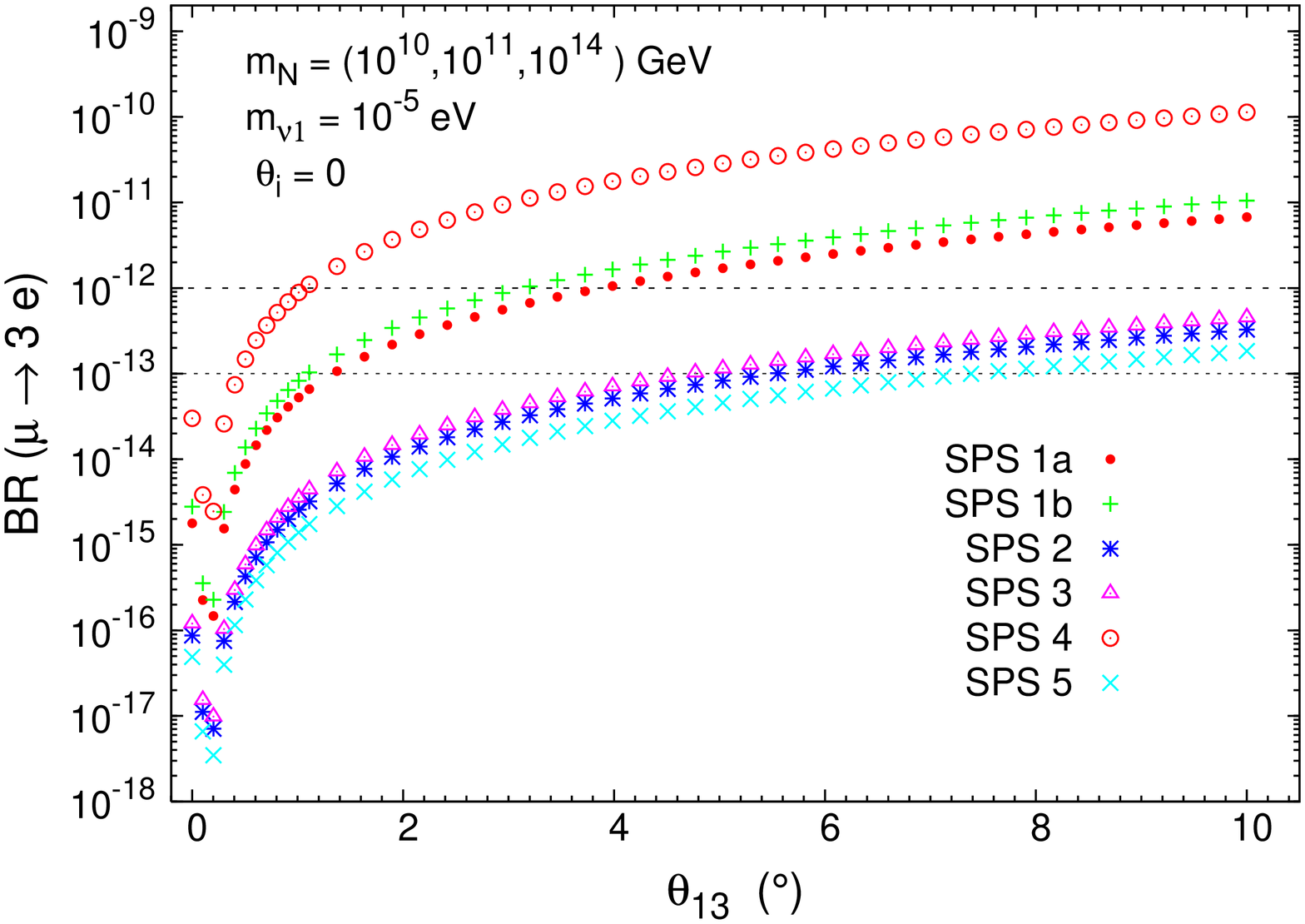,width=50mm,angle=270,clip=}\\
        \psfig{file=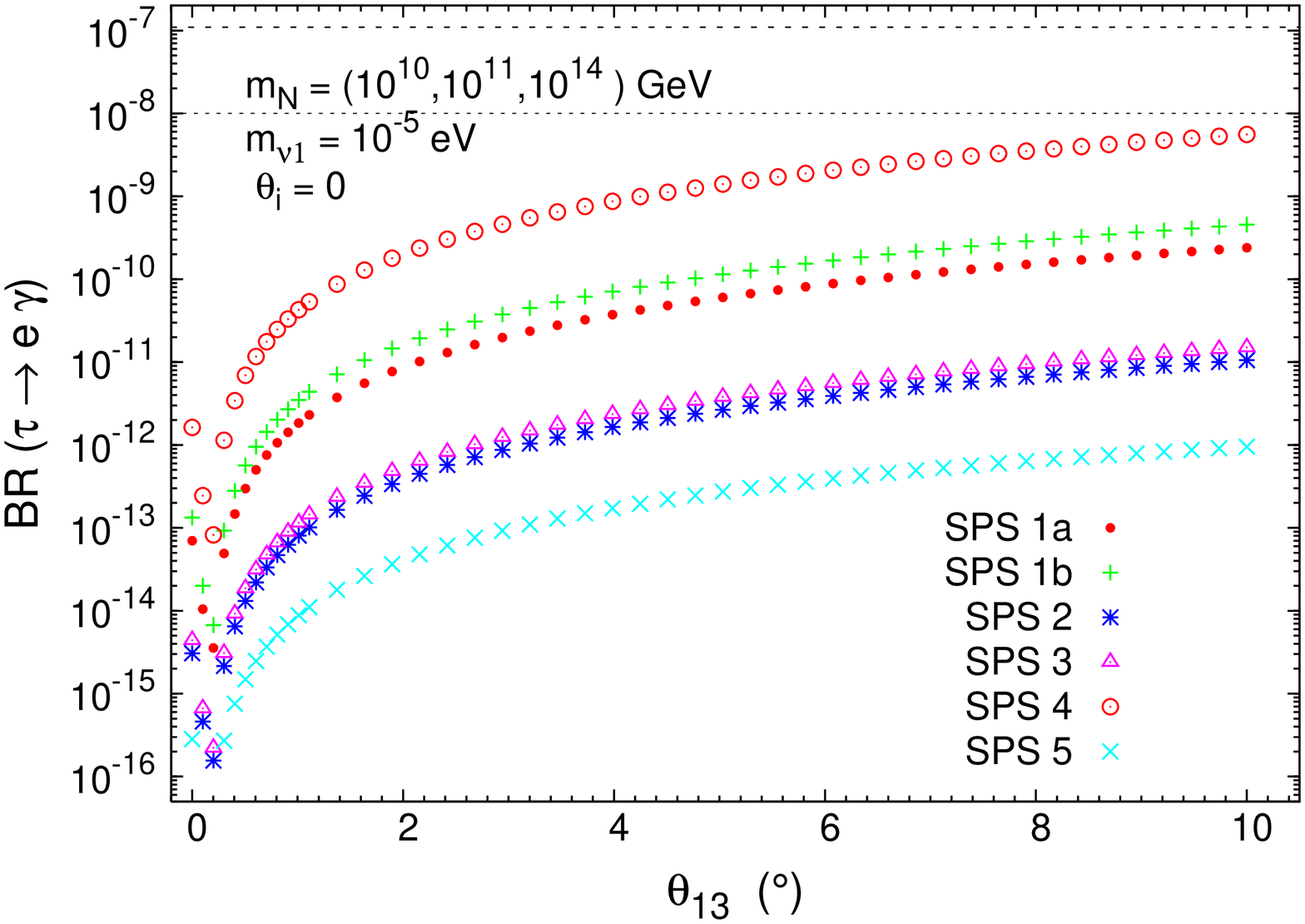,width=50mm,angle=270,clip=} &
        \psfig{file=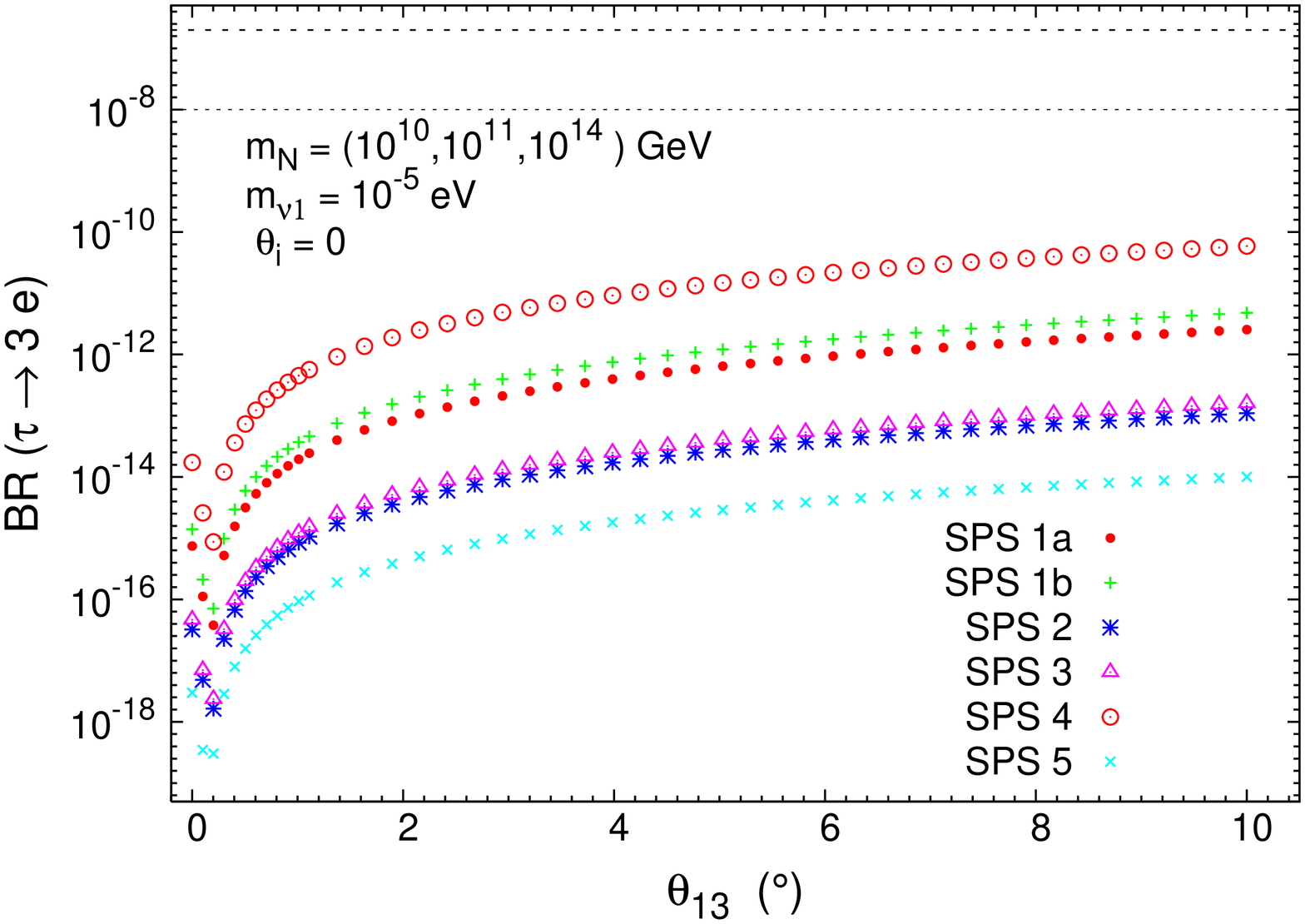,width=50mm,angle=270,clip=} 
   \end{tabular}
    \caption{BR($\mu \to e\, \gamma$) and BR($\mu \to
    3\,e $) as a function of $\theta_{13}$ (in degrees), for SPS 1a
    (dots), 1b (crosses), 2 (asterisks), 3 (triangles), 4 (circles) and 5
    (times). A dashed (dotted)
    horizontal line denotes the present experimental bound (future
    sensitivity).} 
    \label{fig:SPS:t13:ad}
  \end{center}
\end{figure*}

For $R = 1$, the predictions of the BRs as functions of 
$\theta_{13}$, valid within the experimentally allowed range of $\theta_{13}$, $0^\circ \leq
\theta_{13} \lesssim 10^\circ$, are illustrated in Fig.~\ref{fig:SPS:t13:ad}. 
In this figure we also include the present and
future experimental sensitivities for all channels. We clearly see 
that the BRs of  $\mu \to e \gamma$, $\mu \to 3 e$,
$\tau \to e \gamma$ and $\tau \to 3 e$ are extremely sensitive to $\theta_{13}$, with
their predicted rates varying many orders of magnitude along the explored $\theta_{13}$ 
interval. In the  case of $\mu \to e \gamma$ this strong sensitivity was previously 
pointed out in Ref.~\cite{Masiero:2004js}. The other LFV channels, $\tau \to \mu \gamma$ and 
$\tau \to 3 \mu$ (not displayed here), are nearly insensitive to this
parameter. The most important conclusion from Fig.~\ref{fig:SPS:t13:ad} is
that, for this choice of parameters, the predicted BRs for both 
muon decay
channels, $\mu \to e \gamma$ and
$\mu \to 3 e$, are clearly within the present 
experimental reach for
several  of the studied SPS points. The most stringent channel is manifestly $\mu \to e \gamma$ where the
predited BRs for all the SPS points are clearly above the present experimental bound 
for $\theta_{13} \gtrsim 5^\circ$. With the expected improvement in the experimental
sensitivity to this channel, this would happen for $\theta_{13} \gtrsim 1^\circ$.

\vspace{1cm}

In addition to the generation of small neutrino masses, the seesaw
mechanism offers the interesting possibility of 
baryogenesis via leptogenesis~\cite{Fukugita:1986hr}. Thermal leptogenesis 
is an attractive and minimal mechanism to produce a successfull BAU
which is compatible with present data, $n_\mathrm{B} /n_\gamma 
\,\approx\, (6.10\,\pm\,0.21)\,\times\,10^{-10}$ \cite{Spergel:2006hy}. 
In the SUSY version of the seesaw mechanism, it can
be successfully implemented provided that the 
following conditions can be satisfied. Firstly, Big Bang Nucleosynthesis gravitino 
problems have to be avoided, which is possible, for instance, for 
sufficiently heavy gravitinos. Since we consider the gravitino mass 
as a free parameter, this condition can be easily achieved. In any case, further 
bounds on the reheat temperature $T_\mathrm{RH}$ still arise from 
decays of gravitinos into Lightest Supersymmetric Particles (LSPs). In the
case of heavy gravitinos, and neutralino LSPs masses in the range 100-150
GeV (which is the case of the present work), one obtains 
$T_\mathrm{RH} \lesssim 2 \times 10^{10}$ GeV.
In the presence of these constraints on 
$T_\mathrm{RH}$, the favoured region by thermal leptogenesis 
corresponds to small (but non-vanishing) complex $R$-matrix angles 
$\theta_i$. For vanishing $U_\text{MNS}$ CP phases the constraints on $R$ are
basically $|\theta_2|,|\theta_3|  
\lesssim 1 \, \mbox{rad}$ (mod $\pi$). 
Thermal leptogenesis also
constrains $m_{N_1}$ to be roughly in the range $[10^9\:
\mbox{GeV},10 \times T_\mathrm{RH}]$. In the present work, we require the BAU to
be within the interval $[10^{-10},10^{-9}]$, which contains
the WMAP range, and choose the value of $m_{N_1} =
10^{10}$ GeV in some of our plots.
Similar studies of the constraints from leptogenesis on LFV rates have been
done in~\cite{Petcov:2005jh}.

Concerning the EDMs, which are clearly non-vanishing in the presence
of complex $\theta_i$, we have checked that all the predicted values for 
the electron, muon
and tau EDMs are well below the experimental bounds.  
In the following we therefore focus on complex but small $\theta_2$ values,  leading to
favourable BAU,  and study its effects
on the sensitivity to $\theta_{13}$. Similar results are obtained for
$\theta_3$, but for shortness are not shown here. 

\begin{figure*}[t]
  \begin{center} \hspace*{-10mm}
    \begin{tabular}{cc}
      \psfig{file=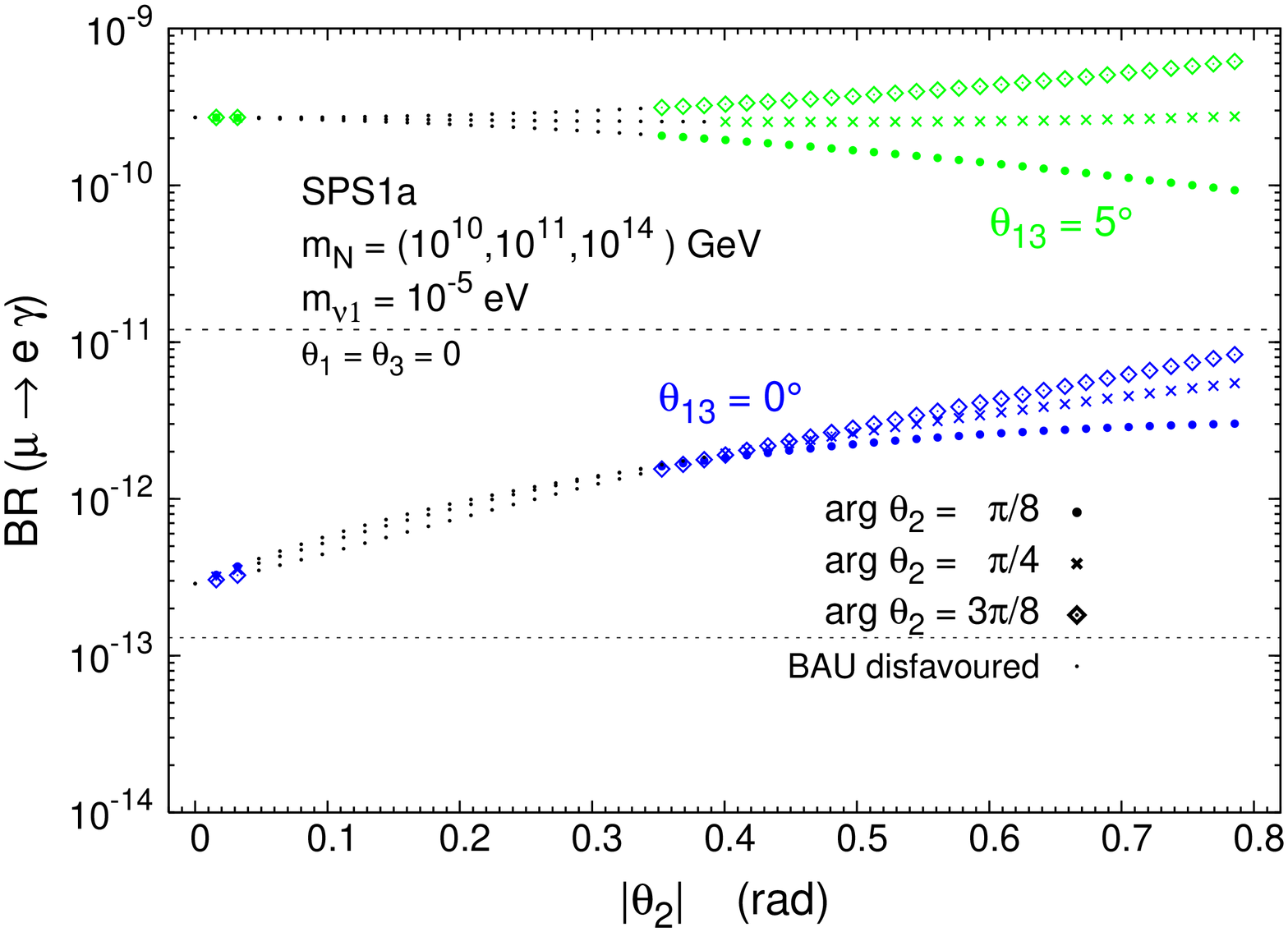,width=55mm,angle=270,clip=}
      &
      \psfig{file=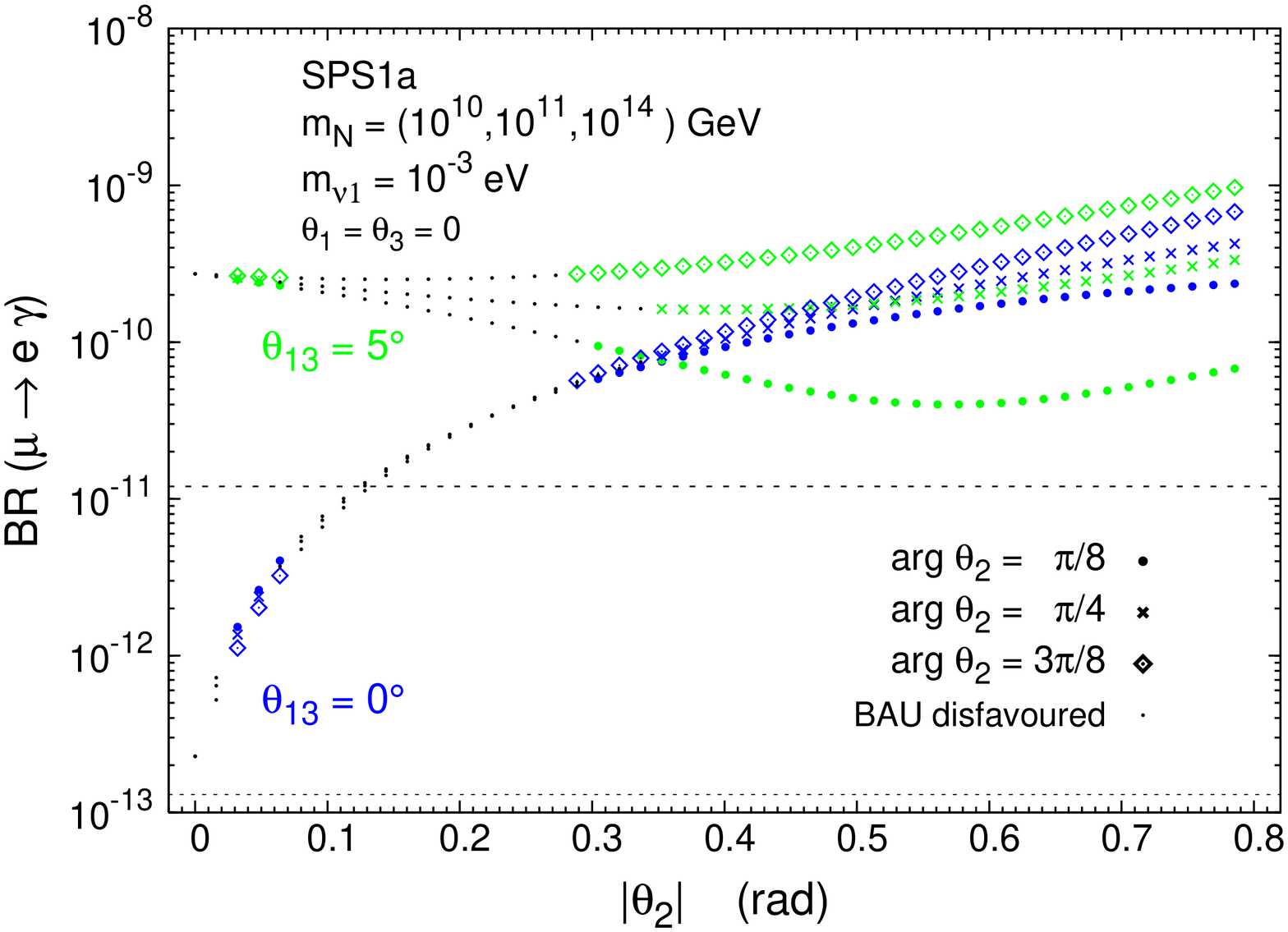,width=55mm,angle=270,clip=}
    \end{tabular}
    \caption{BR($\mu \to e\, \gamma$) as a function
      of $|\theta_2|$, for $\arg
      \theta_2\,=\,\{\pi/8\,,\,\pi/4\,,\,3\pi/8\}$ (dots, times,
      diamonds, respectively) and
      $\theta_{13}=0^\circ$, 5$^\circ$ (blue/darker, green/lighter lines).
      We take $m_{\nu_1}\,=\,10^{-5}$
      ($10^{-3}$) eV, on the left (right) panel. 
      In all cases black dots
      represent points associated with a disfavoured BAU scenario and 
      a dashed (dotted) horizontal line denotes the present 
      experimental bound (future sensitivity).} 
    \label{fig:modt2:argt2:1214}
  \end{center}
\end{figure*}

Fig.~\ref{fig:modt2:argt2:1214} shows the dependence of 
the most sensitive BR to $\theta_{13}$, BR$(\mu \to e\, \gamma)$, on
$|\theta_2|$. We consider two
particular values of $\theta_{13}$,
$\theta_{13}=0^\circ\,,5^\circ$ and choose SPS 1a. Motivated from the thermal leptogenesis favoured $\theta_2$-regions \cite{Antusch:2006vw}, we take $0 \,\lesssim \, |\theta_2|
\,\lesssim \, \pi/4$, with $\arg \theta_2 \,=\,\{\pi/8\,,\,\pi/4\,,\,3\pi/8\}$. 
We display the numerical results,
considering $m_{\nu_1}\,=\,10^{-5}$ eV and $m_{\nu_1}\,=\,10^{-3}$~eV, 
while for the heavy neutrino masses we 
take $m_{N}\, =\, (10^{10},\,10^{11},\,10^{14})$~GeV.
There are several important conclusions to be drawn from
Fig.~\ref{fig:modt2:argt2:1214}. Let us first discuss the case 
$m_{\nu_1}\,=\,10^{-5}$~eV. We note that one can obtain a baryon asymmetry 
in the range $10^{-10}$ to $10^{-9}$
for a considerable region of the analysed $|\theta_2|$ range. 
Notice also that there is a clear separation between the predictions
of $\theta_{13}=0^\circ$ and $\theta_{13}=5^\circ$, with the latter
well above the present experimental bound. This would imply an
experimental impact of $\theta_{13}$, in the sense that the BR predictions 
become potentially
detectable for this non-vanishing $\theta_{13}$ value. 
With the planned
MEG sensitivity~\cite{mue:Ritt}, both cases would be within experimental reach.
However, this statement is strongly dependent on the
assumed parameters, in particular $m_{\nu_1}$.
For instance, a larger value of $m_{\nu_1}=10^{-3}$~eV, illustrated on
the right panel of Fig.~\ref{fig:modt2:argt2:1214}, leads to a very
distinct situation regarding the sensitivity to $\theta_{13}$. 
While for smaller values of $|\theta_2|$
the branching ratio displays a clear sensitivity to having
$\theta_{13}$ equal or different from zero (a separation larger than two orders of
magnitude for $|\theta_2| \lesssim 0.05$), the effect of $\theta_{13}$ is
diluted for increasing values of $|\theta_2|$. 

Let us now address the question of whether a joint measurement of the
BRs and $\theta_{13}$ can shed some light on experimentally unreachable
parameters, like $m_{N_3}$. 
The expected improvement in the experimental sensitivity to the LFV
ratios supports the possibility that 
a BR could be measured in the future, thus
providing the first experimental evidence for new
physics, even before its discovery at the LHC.
The prospects are especially encouraging regarding $\mu \to e\,
\gamma$, where the experimental sensitivity will improve by at least two
orders of magnitude.
Moreover, and given the impressive
effort on experimental neutrino physics, a measurement
of $\theta_{13}$ will likely also occur in the 
future~\cite{theta13_sensitivities}.  
Given that, as previously emphasised, $\mu \to e\,\gamma$ is very
sensitive to $\theta_{13}$, whereas this is not the case for 
BR($\tau \to \mu\,\gamma$), 
and that both BRs display the same approximate behaviour with 
$m_{N_3}$ and $\tan \beta$, we now propose to study the correlation between
these two observables. This optimises the impact of a 
$\theta_{13}$ measurement, since it allows to minimise the uncertainty
introduced from not knowing $\tan \beta$ and $m_{N_3}$, and at the
same time offers a better illustration of the uncertainty associated
with the $R$-matrix angles.
In this case, the correlation of the BRs with respect to $m_{N_3}$
means that, for a fixed set of parameters, varying $m_{N_3}$ implies
that the predicted point 
(BR($\tau \to \mu\,\gamma$),~BR($\mu \to e \, \gamma$)) 
moves along a line with approximately constant slope in the 
BR($\tau \to \mu\,\gamma$)-BR($\mu \to e \, \gamma$) plane.
On the other hand, varying $\theta_{13}$ leads to a 
displacement of the point along the vertical axis.

\begin{figure*}[t]
  \begin{center} \hspace*{-10mm}
	\psfig{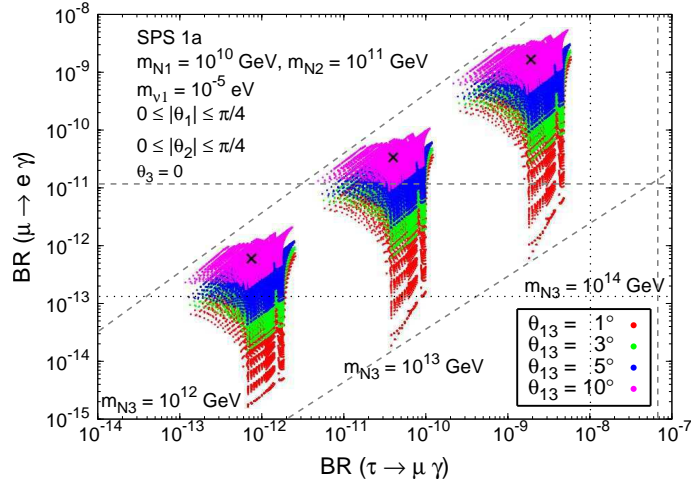} 
    \caption{Correlation between BR($\mu \to e\,\gamma$) and 
      BR($\tau \to \mu\,\gamma$) as a function of $m_{N_3}$, for SPS
      1a. The areas displayed represent the scan over $\theta_i$ 
      as given in Eq.~(\ref{doubleBR:input}). From bottom to top, 
      the coloured regions correspond to 
      $\theta_{13}=1^\circ$, $3^\circ$, $5^\circ$ and $10^\circ$ (red,
      green, blue and pink, respectively). Horizontal and vertical 
      dashed (dotted) lines denote the experimental bounds (future
      sensitivities). } 
    \label{fig:doubleBR}
  \end{center}
\end{figure*}

In Fig.~\ref{fig:doubleBR}, we illustrate this correlation for SPS
1a, choosing distinct values of the heaviest neutrino mass, and
we scan over the BAU-enabling $R$-matrix angles (setting $\theta_3$ to
zero) as 
\begin{align}\label{doubleBR:input}
& 0\, \lesssim \,|\theta_1|\,\lesssim \, \pi/4 \,, \quad 
-\pi/4\, \lesssim \,\arg \theta_1\,\lesssim \, \pi/4 \,, \nonumber \\
& 0\, \lesssim \,|\theta_2|\,\lesssim \, \pi/4 \,, \quad 
\quad \,\,\,\,\,\,
0\, \lesssim \,\arg \theta_2\,\lesssim \, \pi/4 \,, \nonumber \\
& m_{N_3}\,=\,10^{12}\,,\,10^{13}\,,\,10^{14}\,\text{GeV}\,.
\end{align} 
 We consider the following values,
$\theta_{13}=1^\circ$, $3^\circ$, $5^\circ$ and $10^\circ$, and only
include in the plot the BR predictions which allow for a favourable BAU. 
Other SPS points have also been considered but they are not shown here for
brevity (see~\cite{Antusch:2006vw}).
We clearly observe in Fig.~\ref{fig:doubleBR} that  
for a fixed value of $m_{N_3}$, and for a given value of $\theta_{13}$, the
dispersion arising from a $\theta_1$ and $\theta_2$ variation produces a small
area rather than a point in the 
BR($\tau \to\mu\,\gamma$)-BR($\mu \to e \, \gamma$) plane.
The dispersion along the BR($\tau \to \mu\,\gamma$) axis is of
approximately one order of magnitude for all $\theta_{13}$. 
In contrast, the dispersion along the BR($\mu \to e\,\gamma$) axis
increases with decreasing $\theta_{13}$,
ranging from 
an order of magnitude for $\theta_{13}=10^\circ$,
to over three orders of magnitude for the case of small $\theta_{13}$
($1^\circ$). 
From Fig.~\ref{fig:doubleBR} 
we can also infer that other choices of $m_{N_3}$ (for $\theta_{13}
\in [1^\circ, 10^\circ]$) would lead to BR
predictions which would roughly lie within the diagonal lines depicted
in the plot. Comparing
these predictions for the shaded areas along the expected diagonal
``corridor'', with the allowed experimental region, allows to conclude
about the impact of a $\theta_{13}$ measurement on the allowed/excluded 
$m_{N_3}$ values.
The most important conclusion from Fig.~\ref{fig:doubleBR} is that for
SPS~1a, and for the parameter space defined in Eq.~(\ref{doubleBR:input}), 
an hypothetical $\theta_{13}$ measurement larger than $1^\circ$, together 
with the present experimental bound on the BR($\mu \to e\,\gamma$),
will have the impact of excluding values of $m_{N_3} \gtrsim 10^{14}$
GeV. Moreover, with the planned MEG
sensitivity, the same $\theta_{13}$ measurement can further constrain 
$m_{N_3} \lesssim 3\times 10^{12}$~GeV.
The impact of any other $\theta_{13}$ measurement can be analogously
extracted from Fig.~\ref{fig:doubleBR}.

As a final comment let us add that, remarkably, 
within a particular SUSY scenario and scanning over specific $\theta_1$ and $\theta_2$ BAU-enabling ranges for various 
values of $\theta_{13}$, the comparison of the
theoretical predictions for BR($\mu \to e\,\gamma$) and 
BR($\tau \to \mu\,\gamma$) with the present experimental bounds allows 
to set $\theta_{13}$-dependent upper bounds on $m_{N_3}$. 
Together with the indirect lower bound arising from leptogenesis 
considerations, this clearly provides interesting hints on the value of the 
seesaw parameter $m_{N_3}$.
With the planned future sensitivities, these bounds would further improve by
approximately one order of magnitude.
Ultimately, a joint measurement of the LFV branching ratios, 
$\theta_{13}$ and the sparticle spectrum would be a powerful tool for 
shedding some light on otherwise unreachable SUSY seesaw parameters.
It is clear from this study that the connection between LFV and neutrino
physics will play a relevant role for the searches of new physics.

\section*{Acknowledgements}

M.J. Herrero would like to thank Alberto Lusiani for the invitation to participate
in this interesting and fruitful conference. She also acknowledges project
FPA2003-04597 of Spanish MEC for finantial support.

\end{document}